\documentclass[runningheads]{llncs}
\usepackage{geometry}
\usepackage{microtype}
\usepackage{graphicx}
\usepackage[svgnames,pdf]{pstricks}
\usepackage{pstricks}
\usepackage{pst-grad}
\usepackage{pst-plot}
\usepackage{tkz-euclide}
\usepackage{pgfplots}
\usepackage{pgfplotstable}
\usepackage{tkz-euclide}
\usepackage{amsmath}
\usepackage{amssymb}
\usepackage{caption}
\usepackage{subcaption}
\usepackage{graphicx}
\usepackage{xcolor}
\usepackage[utf8]{inputenc}

\usepgfplotslibrary{fillbetween}
\usetikzlibrary{intersections}
\usetikzlibrary{calc}
\pgfplotsset{compat=1.14}

\definecolor{lightcandy}{RGB}{170, 5, 5}
\definecolor{darkcandy}{RGB}{106, 12, 11}
\definecolor{lightblue}{RGB}{103, 199, 235}
\definecolor{darkblue}{RGB}{68, 123, 190}
\definecolor{darkgold}{RGB}{185, 125, 16}
\definecolor{lightgold}{RGB}{251, 202, 3}

\DeclareMathOperator{\Ima}{\text{Im}}
\DeclareMathOperator{\SP}{\text{span}}
\DeclareMathOperator{\rank}{\text{rank}}
\DeclareMathOperator{\Cech}{\text{\v{C}ech}}

\DeclareMathOperator{\Rips}{\text{Rips}}
\DeclareMathOperator{\VorBall}{\text{VorBall}}
\DeclareMathOperator{\Vor}{\text{Vor}}
\DeclareMathOperator{\Del}{\text{Del}}

\DeclareMathOperator{\Wit}{\text{Wit}}
\usepackage[hidelinks]{hyperref}
\hypersetup{
    pdftitle={Persistent Homology as Stopping-Criterion for Voronoi Interpolation},    
    pdfauthor={Luciano Melodia},     
    pdfsubject={Topological Stopping-Criterion},   
    pdfcreator={Luciano Melodia},   
    pdfproducer={Luciano Melodia},  
    pdfkeywords={Persistent Homology} {Interpolation} {Voronoi Triangulation}, 
    pdfnewwindow=true,
    colorlinks,
  citecolor=black,
  filecolor=black,
  linkcolor=black,
  urlcolor=blue
}
\begin{document}
\title{Persistent Homology as Stopping-Criterion for Voronoi Interpolation}
\titlerunning{Topological Stopping for Voronoi Interpolation}
%
\author{Luciano Melodia\orcidID{0000-0002-7584-7287} \\ \and
Richard Lenz\orcidID{0000-0003-1551-4824}}
\authorrunning{L. Melodia \and R. Lenz}
%
\institute{
Chair of Computer Science 6\\
Friedrich-Alexander University Erlangen-Nürnberg\\
91058 Erlangen, Deutschland\\
\email{\{luciano.melodia,richard.lenz\}@fau.de}}
\index{Melodia, Luciano}
\index{Lenz, Richard}
\maketitle              
\begin{abstract}
In this study the Voronoi interpolation is used to interpolate a set of points drawn from a topological space with higher homology groups on its filtration. The technique is based on Voronoi tessellation, which induces a natural dual map to the Delaunay triangulation. Advantage is taken from this fact calculating the persistent homology on it after each iteration to capture the changing topology of the data. The boundary points are identified as critical. The Bottleneck and Wasserstein distance serve as a measure of quality between the original point set and the interpolation. If the norm of two distances exceeds a heuristically determined threshold, the algorithm terminates. We give the theoretical basis for this approach and justify its validity with numerical experiments.

\keywords{Interpolation \and Persistent homology \and Voronoi triangulation}
\end{abstract}

\section{Introduction}
\label{introduction}
Most interpolation techniques ignore global properties of the underlying topological space of a set of points. The topology of an augmented point set depends on the choice of interpoland. However, it does not depend on the topological structure of the data set. The Voronoi interpolation is a technique considering these issues \cite{Bobach2009}. The algorithm has been invented by Sibson \cite{Sibson}. Using Voronoi triangulation to determine the position of a new point respects the topology in terms of simple-homotopy equivalence. For this an implicit restriction to a closed subset of the embedded space is used, see Fig. \ref{fig:voronoi}. The closure of this subset depends on the metric, in Euclidean space it is flat. This restriction, also called \textit{clipping}, leads to varying results for interpolation according to the choice of clip. The clip does not represent the intrinsic geometry nor the topology of the data, but that of the surrounding space. This leads to artifacts during interpolation.

Persistent homology encodes the topological properties and can be calculated in high dimensions. Thus, it is used as indicator for such artifacts \cite{ZomorodianC05}. In particular, this measurement of topological properties behaves stable, i.e. small changes in the coordinate function value also cause small changes in persistent homology \cite{CerriL16}. Efficient data structures and algorithms have been designed to compute persistent homology \cite{ZomorodianC05,Zomorodian10} and scalable ways to compare persistence diagrams using the Wasserstein metric have been developed \cite{CarriereCO17}. This paper uses persistent homology to decide whether a topological change occurs or not.

Up to this point it is an open problem to detect these errors and to terminate the algorithm in time. Our contribution to a solution is divided into three parts:
\begin{itemize}
  \item We introduce persistent homology as a stopping-criterion for interpolation methods. The distance between two persistence diagrams is an indicator of topological changes during augmentation.
  \item We cover the connection of the Voronoi tessellation to the Delaunay triangulation via duality. It is shown that the Delaunay complex is simple-homotopy equivalent to the \v{C}ech complex. We further show that the Delaunay complex is sufficient to compute persistence diagrams.
  \item We investigate the method on a signature data set. It provides interesting and visually interpretable topological features due to the topography of letters. Higher homology groups such as $H_1$ and $H_2$ may appear on the filtration of a signature. This often represents an insurmountable hurdle for other interpolation techniques.
\end{itemize}

\section{Simplicial Complexes and Filtrations}
\label{sec:simplex}
Taking into account the topology of data is beneficial for interpolation, due to the assumption that the point set lies on a topological or even smooth manifold, having a family of smooth coordinate systems to describe it. Another hypothesis says, that the mutual arrangement of every dataset forms some `shape' \cite{ZomorodianC05}, which characterizes the manifold. If the point set changes its shape, it is no longer identifiable with this manifold.

Embedded simplicial complexes, build out of a set of simplices, are suitable objects to detect such shapes, by computing their homology groups. Simplices, denoted by $\sigma$, are the permuted span of the set $X = \{x_0, x_1, \ldots, x_k\} \subset \mathbb{R}^d$ with $k+1$ points, which are not contained in any affine subspace of dimension smaller than $k$ \cite{ParzanchevskiR17}. A simplex forms the convex hull
\begin{equation}
\label{simplex}
\sigma := \left\{ x \in X \; \bigg\rvert \; \sum_{i=0}^{k} \lambda_i x_i \; \text{with} \; \sum_{i=0}^{k} \lambda_i = 1 \; \text{and} \; \lambda_i \geq 0 \right\}.
\end{equation}
Simplices are well-defined embeddings of polyhedra. `Gluing' simplices together at their \textit{faces}, we can construct simplicial complexes out of them. Faces are meant to be $h$-dimensional simplices or $h$-simplices. Informally, the gluing creates a series of $k$-simplices, which are connected by $h$-simplices, that satisfy $h < k$. A finite simplicial complex denoted by $K$ and embedded into Euclidean space is a finite set of simplices with the properties, that each face of a simplex of $K$ is again a simplex of $K$ and the intersection of two simplices is either empty or a common face of both \cite{ParzanchevskiR17}.

We want to take into account the systematic development of a simplicial complex upon a point cloud. This is called filtration and it is the decomposition of a finite simplicial complex $K$ into a nested sequence of sub-complexes, starting with the empty set \cite{ChazalGLM15}:
\begin{flalign}
  \label{filtration}
    \emptyset &= K^{0} \subset K^{1} \subset \cdots \subset K^{n} = K, \\
    K^{t+1} &= K^t \cup \sigma^{t+1}, \quad \text{for} \; t \in \{0,\ldots, n-1\}.
\end{flalign}
In practice a parameter $r$ is fixed to determine the step size of the nested complexes. This can be thought as a `lens' zooming into a certain `granularity' of the filtration. In the following, we present four different simplicial complexes and their theoretical connection.

\subsection{\v{C}ech Complex}
Let the radius $r \geq 0$ be a real number and $B(x,r) = \{y \in \mathbb{R}^d \; | \; ||x-y|| \leq r\}$ the closed ball centered around the point $x \in X \subseteq \mathbb{R}^d$. The \v{C}ech complex for a finite set of points $X$ is defined as
\begin{equation}
  \Cech(X,r) = \{ U \subseteq X  \; \rvert \; \bigcap_{x\in U}B(x,r) \neq \emptyset \}.
\end{equation}
By $||\cdot||$ we denote consequently the $L^2$-norm. In terms of abstract simplicial complexes (see sec. \ref{simcoll}), the \v{C}ech complex is the full abstract simplex spanned over $X$ \cite{bauer2017morse}. According to the Nerve lemma it is homotopy-equivalent to the union of balls $B(X,r) = \bigcup_{x \in U} B(x,r)$ \cite{borsuk}. Spanning the simplicial complex for $r = \sup_{x,y \in U} ||x-y||$, we get the full simplex for the set $U$. For two radii $r_1 < r_2$ we get a nested sequence $\Cech(X,r_1) \subset \Cech(X,r_2)$. This implies that the \v{C}ech complex forms a filtration over $U$ and therefore a filtration over the topological space $X$ if $U = X$ \cite{bauer2017morse}. These properties make the \v{C}ech complex a very precise descriptor of the topology of a point set. The flip side of the coin is that the \v{C}ech complex is not efficiently computable for large point sets. A related complex is therefore presented next, which is slightly easier to compute.

\subsection{Vietoris-Rips Complex}
The Vietoris-Rips complex $\Rips(X,r)$ with vertex set $X$ and distance threshold $r$ is defined as the set
\begin{equation}
  \label{rips}
\Rips(X,r) = \left\{ U \subseteq X \; \bigg\rvert \; ||x-y|| \leq r, \; \text{for all} \; x,y \in U \right\}.
\end{equation}
The Vietoris-Rips complex requires only the comparison of distance measures to be obtained. It spans the same $1$-skeleton as the \v{C}ech complex and fulfills for an embedding into any metric space the following relationship \cite[p.~15]{boissonnat2018geometric}: $\Rips(X,r)\subseteq \Cech(X,r) \subseteq \Rips(X,2r).$ To see this, we choose a simplex $\sigma = \{x_0,x_1,\ldots,x_k\} \in \Rips(X,r)$. The point $x_0 \in \bigcap_{i=0}^{k}B(x_i,r)$ must be within the intersection of closed balls with radius $r$ of all points. Now we choose a $\sigma = \{x_0,x_1,\ldots,x_k\} \in \Cech(X,r)$, then there is a point $y \in \mathbb{R}^d$ within the intersection $y \in \bigcap_{i=0}^{k}B(x_i,r)$, which is the desired condition $d(x_i-y) \leq r$ for any $i = 0,\ldots,k$. Therefore, for all $i,j \in \{0,\ldots,k\}$ the following (in)equality applies: $d(x_i-x_j) \leq 2r$ and $\sigma \in \Rips(X,2r)$.

The calculation time for the Vietoris-Rips complex is better than for the \v{C}ech complex, with a bound of $\mathcal{O}(n^2)$ for $n$ points \cite{Zomorodian10}. As a third complex we introduce the $\alpha$-complex or Delaunay complex, for which the definition of Voronoi cells and balls are prerequisite.
\begin{figure*}[t!]
\centering
\begin{subfigure}{.18\textwidth}
  \centering
  \includegraphics[width=.9\linewidth]{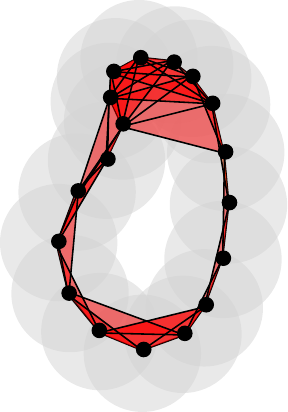}
\end{subfigure}
\hfill
\begin{subfigure}{.18\textwidth}
  \centering
  \includegraphics[width=.9\linewidth]{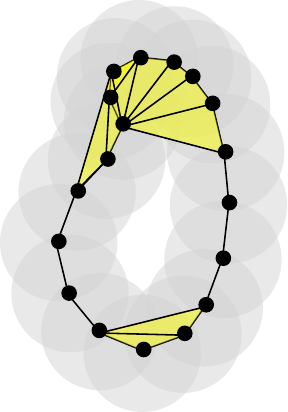}
\end{subfigure}
\hfill
\begin{subfigure}{.18\textwidth}
  \centering
  \includegraphics[width=.9\linewidth]{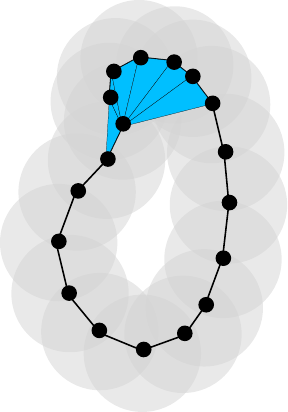}
\end{subfigure}
\hfill
\begin{subfigure}{.18\textwidth}
  \centering
  \includegraphics[width=.9\linewidth]{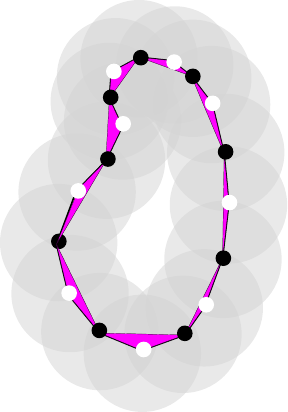}
\end{subfigure}
\hfill
\begin{subfigure}{.18\textwidth}
  \centering
  \includegraphics[width=.9\linewidth]{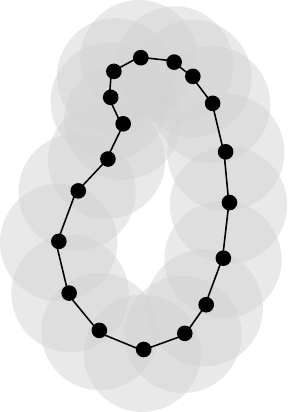}
\end{subfigure}
\caption{Four of five geometric complexes appearing in the collapsing sequence of the \v{C}ech-Delaunay Collapsing Theorem \cite{bauer2017morse}. From \textit{left} to \textit{right}: A high dimensional \v{C}ech complex projected onto the plane, the \v{C}ech-Delaunay complex, the Delaunay complex, the Witness complex, which is an outlier in the row due to the changing shape by different Witness sets (white bullets) and the Wrap complex.}
\label{fig:complexes}
\end{figure*}

\subsection{Delaunay Complex}
If $X \subset \mathbb{R}^d$ is a finite set of points and $x \in X$, then the Voronoi cell or also Voronoi region of a point $x \in X$ is given by
\begin{equation}
\label{voronoicell}
\Vor(x) = \left\{ y \in \mathbb{R}^d \; \bigg\vert \; ||y-x|| \leq ||y-z||, \; \text{for all} \; z \in X \right\}.
\end{equation}
The Voronoi ball of $x$ with respect to $X$ is defined as the intersection of the Voronoi region with the closed ball of given radius around this point, i.e.   $\VorBall(x,r) = B(x,r) \cap \Vor(x)$ \cite{bauer2017morse}.The Delaunay complex on a point set $X$ is defined as
\begin{equation}
\label{delaunay}
  \Del(X,r) = \left\{ U \subseteq X \; \bigg\rvert \; \bigcap_{x\in U} \VorBall(x,r) \neq \emptyset \right\}.
\end{equation}
There is a fundamental connection between the union of all Voronoi balls over $X$ and the Delaunay complex. The idea is to find a \textit{good cover} that does represent the global topology. Taking the topological space $X$ and $U = \bigcup_{i \in I} U_i$ beeing an open cover, we define the Nerve of a cover as its topological structure. Therefore, the empty set $\emptyset \in N(U)$ is part of the Nerve and if $\bigcap_{j\in J} U_j \neq \emptyset$ for a $J \subseteq I$, then $J \in N(U)$. We consider $U$ to be a good cover, if for each $\sigma \subset I$ the set $\bigcap_{i\in \sigma} U_i \neq \emptyset$ is contractible, or in other words if it has the same homotopy type as a point. In this case the Nerve $N(U)$ is homotopy equivalent to $\bigcup_{i \in I} U_i$.

Most interestingly, the Delaunay complex $\text{Del}_r(X)$ of a point set $X$ is isomorphic to the Nerve of the collection of Voronoi balls. To see this, we construct Voronoi regions for two different sets. Thus, we denote the Voronoi region $\Vor(x,r,U)$ of a point within a set $U$. Be $\Vor(x,r,U) \subseteq \Vor(x,r,V)$ for each open set $U \subseteq V \subseteq X$ and all $x \in X$. We obtain the largest Voronoi ball for $U = \emptyset$ and the smallest Voronoi ball for $U = X$. In the first case each region is a ball with radius $r$ and in the second case the Voronoi balls form a convex decomposition of the union of balls. We select a subset $U$ and restrict the Delaunay complex to it by taking into account only the Voronoi balls around the points in $U$. It is called selective Delaunay complex and contains the Delaunay and \v{Cech} complex in its extremal cases:
\begin{equation}
\Del(X,r,U) = \left\{V \subseteq X \; \bigg\vert \; \bigcap_{x \in V} \VorBall(x,r,U) \neq \emptyset \right\}.
\end{equation}
Since the union of open balls does not depend on $U$, the Nerve lemma implies, that for a given set of points $X$ and a radius $r$ all selective Delaunay complexes have the same homotopy type. This also results in $\Del(X,r,V) \subseteq \Del(X,s,U)$ for all $r \leq s$ and $U \subseteq V$. The proof has been given first by \cite[§3.4]{bauer2017morse}.

\subsection{Witness Complex}
Through the restriction of the faces to randomly chosen subsets of the point cloud the filtration is carried out on a scalable complex, which is suitable for large point sets. We call these subsets Witnesses $W \subset \mathbb{R}^d$ and $L \subset \mathbb{R}^d$ landmarks. The landmarks can be part of the Witnesses $L \subseteq W$, but do not have to. Then $\sigma$ is a simplex with vertices in $L$ and some points $w \in W$. We say that $w$ is Witnessed by $\sigma$ if $||w-p|| \leq ||w-q||,$ for all $p \in \sigma$ and $q \in L \setminus \sigma.$ We further say it is strongly Witnessed by $\sigma$ if $||w-p|| \leq ||w-q||,$ for all $p \in \sigma$ and $q \in L$. The Witness complex $\Wit(L,W)$ consists of all simplices $\sigma$, such that any simplex $\tilde{\sigma} \subseteq \sigma$ has a Witness in $W$ and the strong Witness complex analogously.

The homology groups of the Witness complex depend strongly on the landmarks. In addition to equally distributed initialization, strategies such as sequential \textit{MaxMin} can lead to a more accurate estimate of homology groups \cite{de2004topological}. Its time bound for construction is $\mathcal{O}\left(|W|log|W|+k|W|\right)$ \cite{boissonnat2018geometric}.

\section{Persistent Homology Theory}
\label{sec:simplicialhomology}
We are particularly interested in whether a topological space can be continuously transformed into another. For this purpose its $k$-dimensional `holes' play a central role. Given two topological spaces $M$ and $N$ we say that they have the same \textit{homotopy type}, if there exists a continuous map $h: M \times I \rightarrow N$, which deforms $M$ over some time interval $I$ into $N$. But it is very difficult to obtain homotopies. An algebraic way to compute something strongly related is homology. The connection to homotopy is established by the Hurewicz Theorem. It says, that given $\pi_k(x,X)$, the $k$-th homotopy group of a topological space $X$ in a point $x \in X$, there exists a homomorphism $h: \pi_k(x,X) \rightarrow H_k(x,X)$ into the $k$-th homology group at $x$. It is an isomorphism if $X$ is $(n-1)$-connected and $k \leq n$ when $n \geq 2$ with abelianization for $n = 1$ \cite{hatcher2005algebraic}. In this particular case, we are able to use an easier to calculate invariant to describe the topological space of the data up to homotopy. Further we need to define what a boundary and what a chain is, respectively. We want to describe the boundary of a line segment by its two endpoints, the boundary of a triangle, or $2$-simplex by the union of the edges and the boundary of a tetrahedron, or $3$-simplex by the union of the triangular faces. Furthermore, a boundary itself shall not have a boundary of its own. This implies the equivalence of the property to be boundaryless to the concept of a `loop', i.e. the possibility to return from a starting point to the same point via the $k$-simplices, by not `entering' a simplex twice and not `leaving' a simplex `unentered'.

Let $\sigma^k$ be a $k$-simplex of a simplicial complex $K := K(X)$ over a set of points $X$. Further, let $k \in \mathbb{N}$. The linear combinations of $k$-simplices span a vector space $C_k := C_k(K) = \SP\left(\sigma_1^k, \ldots, \sigma_n^k\right).$ This vector space is called $k$-th chain group of $K$ and contains all linear combinations of $k$-simplices. The coefficients of the group lie in $\mathbb{Z}$ and the group structure is established by $(C_k,+)$, with $e_{C_k}=0$ being the neutral element and addition as group operation. A linear map $\partial:C_k \rightarrow C_{k-1}$ is induced from the $k$-th chain group into the $(k-1)$-th.  The boundary operator $\partial_k(\sigma^k): C_k \rightarrow C_{k-1}$ is defined by
\begin{equation}
  \partial_k(\sigma^k) = \sum_{i=0}^{k} (-1)^i \left(v_0, \ldots, \widehat{v_i},\ldots,v_k\right).
\end{equation}
The vertex set of the $k$-simplex is $v_0,\ldots,v_k$. This group homomorphism contains an alternating sum, thus for each oriented $k$-simplex $(v_0, \ldots, v_k)$ one element $\widehat{v_i}$ is omitted. The boundary operator can be composed $\partial^2 := \partial \circ \partial$. We observe, that every chain, which is a boundary of higher-dimensional chains, is boundaryless. An even composition of boundary maps is zero $\partial^{2\mathbb{Z}} = 0$ \cite{hatcher2005algebraic}.

The kernel of $(C_k,+)$ is the collection of elements from the $k$-th chain group mapped by the boundary operator to the neutral element of $(k-1)$-th: $\ker \partial_k = \partial_k^{-1}(e_{C_{k-1}}) = \{\sigma^k \in C_k \; \vert \; \partial_k(\sigma^k)= e_{C_{k-1}}\}$. A cycle should be defined by having no boundary. From this we get a group of $k$-cycles, denoted by $Z_k$, which is defined as the kernel of the $k$-th boundary operator $Z_k := \ker \partial_k \subseteq C_k$. Every $k$-simplex mapped to zero by the boundary operator is considered to be a cycle and the collection of cycles is the group of $k$-cycles $Z_k$. The $k$-boundaries are therefore $B_k = \Ima \partial_{k+1} \subset Z_k$. The $k$-th homology group $H_k$ is the quotient
\begin{equation}
  H_k := Z_k / B_k = \ker \partial_k / \Ima \partial_{k+1}.
\end{equation}
We compute the $k$-th Betti numbers by the rank of this vector space, i.e. $\beta_k = \rank H_k$. In a certain sense the Betti numbers count the amount of holes in a topological space, i.e. $\beta_0$ counts the connected components, $\beta_1$ the tunnels, $\beta_2$ voids and so forth. Using Betti numbers, the homology groups can be tracked along the filtration, representing the `birth' and `death' of homology classes. The filtration of a simplicial complex defines a sequence of homology groups connected by homomorphisms for each dimension. The $k$-th homology group over a simplicial complex $K_r$ with parameter $r$ is denoted by $H_k^r = H_k(K_r)$. This gives a group homomorphism $g^{r,r+1}_k: H_k^r \rightarrow H_k^{r+1}$ and the sequence \cite{edelsbrunner2008persistent}:
\begin{equation}
0 = H_k^0 \xrightarrow{g^{0,1}_k} H_k^1 \xrightarrow{g^{1,2}_k} \cdots \xrightarrow{g^{n,r}_k} H_k^r \xrightarrow{g^{r,r+1}_k} H_k^{r+1} = 0.
\end{equation}
The image $\Ima g^{r,r+1}_k$ consists of all $k$-dimensional homology classes which are born in the $K_r$-complex or appear before and die after $K_{r+1}$. The dimension $k$ persistent homology group is the image of the homomorphisms $H_k^{n,r} = \Ima g_k^{n,r}$, for $0\leq n \leq r \leq r+1$ \cite{edelsbrunner2008persistent}. For each dimension there is an index pair $n \leq r$. Tracking the homology classes in this way yields a multi set, as elements from one homology group can appear and vanish several times for a certain parametrization. Thus, we get the following multiplicity:
\begin{equation}
\mu^{n,r}_k = \underbrace{(\beta_k^{n,r-1} - \beta_k^{n,r})}_{\text{Birth in $K_{r-1}$, death at $K_r$.}} - \underbrace{(\beta_k^{n-1,r-1}-\beta_k^{n-1,r})}_{\text{Birth before $K_r$, death at $K_r$.}}
\end{equation}
The first difference counts the homology classes born in $K_{r-1}$ and dying when $K_r$ is entered. The second difference counts the homology classes born before $K_{r-1}$ and dying by entering $K_r$. It follows that $\mu^{n,r}_k$ counts the $k$-dimensional homology classes born in $K_n$ and dying in $K_r$ \cite{edelsbrunner2008persistent}.

The persistence diagram for the $k$-th dimension, denoted as $\mathcal{P}^{(\text{dim} k)}_{K}$, is the set of points $(n,r) \in \mathbb{\bar{R}}^2$ with $\mu^{n,r}_k = 1$ where $\bar{\mathbb{R}} := \mathbb{R} \cup +\infty$. We define the general persistent diagram as the disjoint union of all $k$-dimensional persistence diagrams $\mathcal{P}_K = \bigsqcup_{k \in \mathbb{Z}} \mathcal{P}^{(\text{dim} k)}_{K}.$ In this paper we consider $H_0, H_1$ and $H_2$. We now introduce distances for comparison of persistence diagrams. In particular, it is important to resolve the distance between multiplicities in a meaningful way. Note that they are only defined for $n < r$ and that no values appear below the diagonal. This is to be interpreted such that a homology class can't disappear before it arises.

\section{Bottleneck Distance}
Let $X$ be a set of points embedded in Euclidean space and $K_{r}^1, K_{r}^2$ two simplicial complexes forming a filtration over $X$. Both are finite and have in all their sub-level sets homology groups of finite rank. Note, that these groups change due to a finite set of homology-critical values. To define the bottleneck distance we use the $L^\infty$-norm $||x-y||_\infty = \max\left\{|x_1-y_1|,|x_2-y_2|\right\}$ between two points $x = (x_1,x_2)$ and $y = (y_1,y_2)$ for $x \in \mathcal{P}_{K^1}$ and $y \in \mathcal{P}_{K^2}$. By convention, it is assumed that if $x_2=y_2=+\infty$, then $||x-y||_\infty = |x_1-y_1|$. If $\mathcal{P}_{K^1}$ and $\mathcal{P}_{K^2}$ are two persistence diagrams and $x:=(x_1,x_2) \in \mathcal{P}_{K^1}$ and $y := (y_1,y_2)\in \mathcal{P}_{K^2}$, respectively, their Bottleneck distance is defined as
\begin{equation}
d_\text{B}(\mathcal{P}_{K^1},\mathcal{P}_{K^2})=\inf_{\varphi} \; \sup_{x\in \mathcal{P}_{K^1}}\;||x-\varphi(x)||_\infty,
\end{equation}
where $\varphi$ is the set of all bijections from the multi set $\mathcal{P}_{K^1}$ to $\mathcal{P}_{K^2}$ \cite{boissonnat2018geometric}.

\subsection{Bottleneck Stability}
We consider a smooth function $f: \mathbb{R} \rightarrow \mathbb{R}$ as a working example. A point $x \in \mathbb{R}$ of this function is called critical and $f(x)$ is called critical value of $f$ if $df_x = 0$. The critical point is also said to be not degenerated if $d^2f_x \neq 0$. $\Ima f(x)$ is a homology critical value, if there is a real number $y$ for which an integer $k$ exists, such that for a sufficiently small $\alpha > 0$ the map $H_k\left(f^{-1}\left((-\infty,y-\alpha]\right)\right)\rightarrow H_k\left(f^{-1}\left((-\infty,y+\alpha]\right)\right)$ is not an isomorphism. We call the function $f$ tame if it has a finite number of homology critical values and the homology group $H_k\left(f^{-1}\left((-\infty,y]\right)\right)$ is finite-dimensional for all $k\in \mathbb{Z}$ and $y \in \mathbb{R}$. A persistence diagram can be generated by pairing the critical values with each other and transferring corresponding points to it.

The Bottleneck distance of the persistence diagram of two tame functions $f,g$ is restricted to a norm between a point and its bijective projection. Therefore, not all points of a multi set can be mapped to the nearest point in another \cite{Cohen-SteinerEH07}. To see this, we consider $f$ to be tame. The Hausdorff distance $d_H(X,Y)$ between two multi sets $X$ and $Y$ is defined by
\begin{align}
\max \left\{\sup_{x \in X} \inf_{y \in Y} ||x-y||_\infty, \sup_{y \in Y} \inf_{x \in X} ||y-x||_\infty \right\}.
\end{align}
From the results of \cite{CerriL16} it is known that the Hausdorff (in)equality $d_H(\mathcal{P}_f,\mathcal{P}_g) \leq ||f-g||_\infty = \alpha$ holds and that there must exist a point $(x_1,x_2) \in \mathcal{P}_f$ which has a maximum distance $\alpha$ to a second point $(y_1,y_2) \in \mathcal{P}_g$. In particular, $(y_1,y_2)$ must be within the square $[x_1 -\alpha, x_1 + \alpha] \times [x_2-\alpha, x_2+\alpha]$. Let $x_1 \leq x_2 \leq x_3 \leq x_4$ be points in the extended plane $\bar{\mathbb{R}}^2$. Further, let $R = [x_1,x_2] \times [y_1,y_2]$ be a square and $R_\alpha = [x_1+\alpha,x_2-\alpha] \times [y_1+\alpha,y_2-\alpha]$ another shrinked square by some parameter $\alpha$. Thus, we yield
\begin{equation}
\label{boxlemma}
\#\left( \mathcal{P}_f \cap R_\alpha \right) \leq \#\left(\mathcal{P}_g \cap R \right).
\end{equation}
We need the inequality to find the smallest $\alpha$ such that squares of side-length $2 \alpha$ centered at the points of one diagram cover all off-diagonal elements of the other diagram, and vice versa with the diagrams exchanged \cite{Cohen-SteinerEH07}. The persistence diagrams $\mathcal{P}_f$ and $\mathcal{P}_g$ satisfy for two tame functions $f,g: X \rightarrow \mathbb{R}$:
\begin{equation}
d_B\left(\mathcal{P}_f,\mathcal{P}_g\right) \leq ||f-g||_\infty.
\end{equation}
We take two points $x = (x_1,x_2), y=(y_1,y_2) \in \mathcal{P}_f$ and look at the infinite norm between them in the persistence diagram of $f$ outside the diagonal $\Delta$. In case that there is no such second point we consider the diagonal itself:
\begin{equation}
\delta_f = \min\left\{ ||x-y||_\infty \; \vert \; \mathcal{P}_f-\Delta \ni x \neq y \in \mathcal{P}_f \right\}.
\end{equation}
We choose a second tame function $g$, which satisfies $||f-g||_\infty \leq \delta_f/2$.
We center a square $R_\alpha(x)$ at $x$ with radius $\alpha = ||f-g||_\infty$. Applying Eq. \ref{boxlemma} yields
\begin{equation}
\mu \leq \#(\mathcal{P}_g \cap R_\alpha(x)) \leq \#(\mathcal{P}_f \cap R(x)_{2\alpha}).
\end{equation}
Since $g$ was chosen in such a way, that $||f-g||_\infty \leq \delta_f / 2$ applies, we conclude that $2 \alpha \leq \delta_f$. Thus, $x$ is the only point of the persistence diagram $\mathcal{P}_f$ that is inside $R_{2\alpha}$ and the multiplicity $\mu$ is equal to $\#(\mathcal{P}_g \cap R(x)_\alpha)$. We can now project all points from $\mathcal{P}_g$ in $R(x)_\alpha$ onto $x$. As $d_H(\mathcal{P}_f, \mathcal{P}_g) \leq \alpha$ holds, the remaining points are mapped to their nearest point on the diagonal.

\section{Wasserstein Distance}
The Wasserstein distance is defined for separable completely metrizable topological spaces. In this particular case between the two persistence diagrams $\mathcal{P}_{K^1}$ and $\mathcal{P}_{K^2}$. The $L^p$-Wasserstein distance $W^p$ is a metric arising from the examination of transport plans between two distributions and is defined for a $p \in [1,\infty)$ as
\begin{equation}
d_{W^p}(\mathcal{P}_{K^1},\mathcal{P}_{K^2}) = \left(\inf_{\varphi} \sum_{x \in \mathcal{P}_{K^1}} ||x-\varphi(x)||^p_\infty\right)^{1/p}.
\end{equation}
Then $\varphi: \mathcal{P}_{K^1}\rightarrow \mathcal{P}_{K^2}$ is within the set of all transportation plans from $\mathcal{P}_{K^1}$ to $\mathcal{P}_{K^2}$ over $\mathcal{P}_{K^1} \times \mathcal{P}_{K^2}$. We use the $L^1$-Wasserstein distance. The Wasserstein distance satisfies the axioms of a metric \cite[p.~77]{villani2008optimal}. The transportation problem can be stated as finding the most economical way to transfer the points from one persistence diagram into another. We assume that these two persistence diagrams are disjoint subsets of $\bar{\mathbb{R}}^2 \times \bar{\mathbb{R}}^2$. The cost of transport is given by $d: \bar{\mathbb{R}}^2 \times \bar{\mathbb{R}}^2 \rightarrow [0,\infty)$, so that $||x-\varphi(x)||$ indicates the length of a path. The transport plan is then a bijection $\varphi: \mathcal{P}_{K^1} \rightarrow \mathcal{P}_{K^2}$ from one persistence diagram to the other. The Wasserstein distance of two persistence diagrams is the optimal cost of all transport plans. Note, that the $L^\infty$-Wasserstein distance is equivalent to the Bottleneck distance, i.e. $d_{B}$ is the limit of $d_{W^p}$ as $p \rightarrow \infty$.

\subsection{Wasserstein Stability}
The distance $d_{W^p}$ is stable in a trianguliable compact metric space, which restricts it to Lipschitz continuous functions for stable results. A function $f: X \rightarrow Y$ is called Lipschitz continuous, if one distance $(X,d_X)$ is bounded by the other $(Y,d_Y)$ times a constant, i.e. $d_Y(f(x_1) - f(x_2)) \leq c \cdot d_X(x_1-x_2)$, for all $x_1, x_2 \in X.$ For two Lipschitz functions $f,g$ constants $b$ and $c$ exist \cite{cohen2010lipschitz}, which depend on $X$ and the Lipschitz constants of $f$ and $g$, such that the $p$-th Wasserstein distance between the two functions satisfies
\begin{equation}
  d_{W^p}(f,g) \leq c \cdot ||f-g||_\infty^{1-b/p}.
\end{equation}
For small enough perturbations of Lipschitz functions their $p$-th Wasserstein distance is bounded by a constant. In Fig. \ref{fig:stopping} the topological development of handwritings through interpolation is visualized. Equally colored lines represent the same user and each line represents a signature. The equally colored lines show very similar behavior and represent the small perturbations, which are caused by the slight change of letter shape when signing multiple times.

\section{The Natural Neighbor Algorithm}
\label{sec:natneighbor}
\begin{figure}[t!]
\centering
\begin{subfigure}{.23\textwidth}
  \centering
  \includegraphics[width=\linewidth]{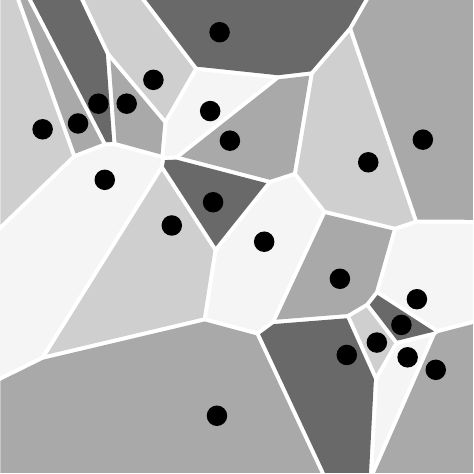}
\end{subfigure}\hfill
\begin{subfigure}{.23\textwidth}
  \centering
  \includegraphics[width=\linewidth]{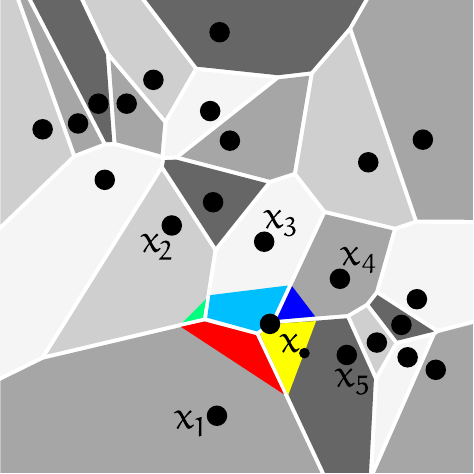}
\end{subfigure}\hfill
\begin{subfigure}{.23\textwidth}
  \centering
  \includegraphics[width=\linewidth]{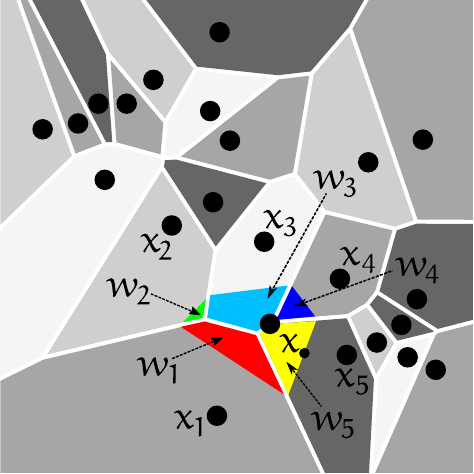}
\end{subfigure}\hfill
\begin{subfigure}{.23\textwidth}
  \centering
  \includegraphics[width=\linewidth]{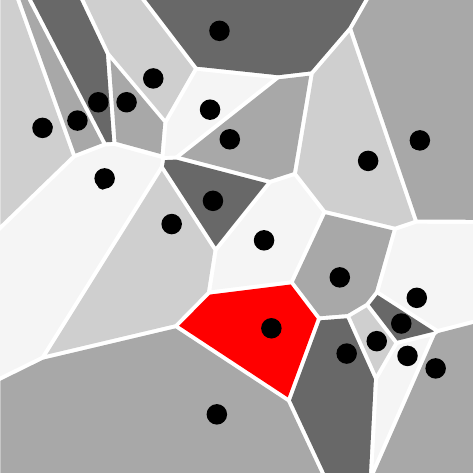}
\end{subfigure}
\caption{From \textit{left} to \textit{right}: Clipped tessellation. For other manifolds the curvature should be considered; Tessellation with added point creates a new Voronoi region \textit{stealing} area from the neighboring regions; The determined weights by the fractional amount of occupied area; Tessellation with added point.}
\label{fig:voronoi}
\end{figure}

The algorithm re-weights the coordinates of a new point in the convex hull of a point cloud by the change of Voronoi regions relative to the Voronoi regions without the additional point, i.e. $\hat{x}_{\bullet} = \sum_{l=1}^{L} \lambda_l x_{\bullet l}$. For a set of points $X \subset \mathbb{R}^d$ distributed over an embedded manifold $M$ natural neighbors behave like a local coordinate system for $M$ with their density increasing \cite{BoissonnatC01}.

The Voronoi tessellation is dual to the Delaunay triangulation, thus we use the latter for our computation \cite[pp.~45-47]{BoissonnatDG18}. This duality gives a bijection between the faces of one complex and the faces of the other, including incidence and reversibility of operations. Both have the same homotopy type. A Voronoi diagram $\text{dgm}_{\Vor}(X)$ is defined by the union of the Voronoi regions $\text{dgm}_{\Vor}(X) := \bigcup_{x \in X} \Vor(x),$ for all $x \in X$ and assigns a polyhedron to each point, see Fig. \ref{fig:voronoi}. This interpolation method generalizes to arbitrary dimensions.

The combinatorial complexity of the Voronoi diagram of $n$ points of $\mathbb{R}^d$ is at most the combinatorial complexity of a polyhedron defined as the intersection of $n$ half-spaces of $\mathbb{R}^{d+1}$. Due to duality the construction of $\text{dgm}_{\Vor}(X)$ takes $\mathcal{O}( n \log n+n^{d/2})$  \cite{BoissonnatDG18}.

\subsection{Voronoi Tessellation}
The Voronoi cells have no common interior, intersect at their boundaries and cover the entire $\mathbb{R}^d$. The resulting polygons can then be divided into Voronoi edges and vertices. The natural neighbors of a point are defined by the points of the neighboring Voronoi polygons \cite{Sibson}.

The natural neighbor is the closest point $x$ to two other points $y$ and $z$ within $X$. To yield the position of the added point we have to calculate the Voronoi diagram of the original signature $\text{dgm}_{\Vor}(X)$ and one with an added point $\text{dgm}_{\Vor}^{\bullet}(X\cup x_{\bullet}) = \bigcup_{x \in X \cup x_\bullet} \Vor(x)$. The latter consists of one Voronoi region more than the primer, see Fig. \ref{fig:voronoi} (a) and (c). This polygon is part of $\text{dgm}_{\Vor}^{\bullet}(X\cup x_{\bullet})$ and contains a certain amount of its `area'.

The Voronoi regions sum up to one $\sum_{l=1}^{L} \lambda_l = 1$. The Voronoi interpolation is repeated until the topological stopping condition is met, measured through the $W^p$-distances of the persistence diagrams $\mathcal{P}_{K}$ and $\mathcal{P}_{K^\bullet}$. The weights of the coordinate representation of $x_{\bullet}$ are determined by the quotient of the `stolen' Voronoi regions and the total `area' of the Voronoi diagram with the additional point according to Eq. \ref{voreq} \cite{Anton2001}. 
\begin{equation}
  \label{voreq}
    \lambda_{l} =
    \begin{cases}
        \frac{\text{vol}\left(\Vor(x_l) \cap \Vor^{\bullet}(x_{\bullet}\right))}{\text{vol}\left(\Vor^{\bullet}(x_{\bullet})\right)} & \text{if } x\geq 1,\\
        0 & \text{otherwise.}
    \end{cases}
\end{equation}
But to what extent are homology groups preserved if persistent homology is computed on the Delaunay triangulation?

\section{The Simplicial Collapse}
\label{simcoll}
The Delaunay triangulation avoids the burden of an additional simplicial structure for persistent homology. We determine how accurate the persistent homology is on this filtration. We use results from simplicial collapse \cite{bauer2017morse}, which show the simple-homotopy equivalence of the \v{C}ech and Delaunay complex among other related simplicial complexes. Simple-homotopy equivalence is stronger than homotopy equivalence. An elementary simplicial collapse determines a strong deformation retraction up to homotopy. Hence, simple-homotopy equivalence implies homotopy equivalence \cite[§2]{cohen2012course}. Under the conditions of the Hurewicz Theorem we can draw conclusions about the homotopy groups of the data manifold.

The simplicial collapse is established using abstract simplicial complexes denoted by $\tilde{K}$. A family of simplices $\sigma$ of a non-empty finite subset of a set $\tilde{K}$ is an abstract simplicial complex if for every set $\sigma'$ in $\sigma$ and every non-empty subset $\sigma'' \subset \sigma'$ the set $\sigma''$ also belongs to $\sigma$. We assume $\sigma$ and $\sigma'$ are two simplices of $\tilde{K}$, such that $\sigma \subset \sigma'$ and $\dim \sigma < \dim \sigma'$. We call the face $\sigma'$ free, if it is a maximal face of $\tilde{K}$ and no other maximal face of $\tilde{K}$ contains $\sigma$. A similar notion to deformation retraction needs to be defined for the investigation of homology groups. This leads to the simplicial collapse $\searrow$ of $\tilde{K}$, which is the removal of all $\sigma''$ simplices, where $\sigma \subseteq \sigma'' \subseteq \sigma'$, with $\sigma$ being a free face. Now we can define the simple-homotopy type based on the concept of simplicial collapse. Intuitively speaking, two simplicial complexes are `combinatorial-equivalent', if it is possible to deform one complex into the other with a finite number of `moves'.  Two abstract simplicial complexes $\tilde{K}$ and $\tilde{G}$ are said to have the same simple-homotopy type, if there exists a finite sequence $\tilde{K}=\tilde{K}_0 \searrow \tilde{K}_1 \searrow \cdots \searrow \tilde{K}_n = \tilde{G}$, with each arrow representing a simplicial collapse or expansion (the inverse operation). If $X$ is a finite set of points in general position in $\mathbb{R}^d$, then
\begin{equation}
\label{theo:collapse}
\text{\v{C}ech}(X,r) \searrow \text{Del\v{C}ech}(X,r) \searrow \text{Del}(X,r) \searrow \text{Wrap}(X,r)
\end{equation}
for all $r \in \mathbb{R}$. For the proof we refer to \cite{bauer2017morse}. The connection in Eq. \ref{theo:collapse} establishes the simple-homotopy equivalence of the \v{C}ech- and Delaunay complex. We deduce, that if the underlying space follows the condition of a Hurewicz isomorphism, all four complexes are suitable for calculating persistent homology as a result of the simplicial collapse up to homotopy equivalence.

\section{Numerical Experiments}
\label{sec:numexp}
\begin{figure}[t!]
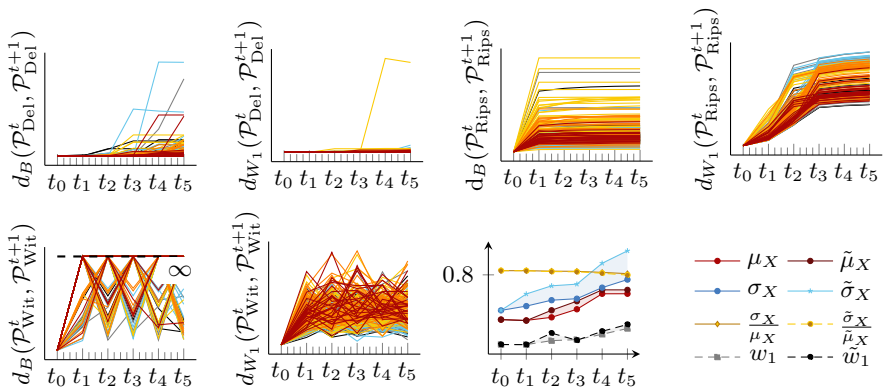

\centering
\begin{tikzpicture}
\begin{axis}[
   width=2cm,
   height=1.5cm,
   scale only axis,
   xtick={1,2,3,4,5,6},
   xticklabels={$t_0$,$t_1$,$t_2$,$t_3$, $t_4$, $t_5$},
   xmajorgrids,
   ylabel={$d_B(\mathcal{P}^{t}_{\Del},\mathcal{P}^{t+1}_{\Del})$},
   ylabel near ticks,
   grid = none,
   minor tick num=3,
   every major grid/.style={darkcandy, opacity=0.5},
   ytick=\empty,
   axis lines*=left,
   cycle list name= color list
]
\input{figure3a.tex}
\end{axis}
\end{tikzpicture}\hfill
\begin{tikzpicture}
\begin{axis}[
   width=2cm,
   height=1.5cm,
   scale only axis,
   xtick={1,2,3,4,5,6},
   xticklabels={$t_0$,$t_1$,$t_2$,$t_3$, $t_4$, $t_5$},
   ylabel={$d_{W_1}(\mathcal{P}^{t}_{\Del},\mathcal{P}^{t+1}_{\Del})$},
   ylabel near ticks,
   ytick=\empty,
   axis lines*=left,
   cycle list name= color list,
   grid = none,
   minor tick num=3,
   every major grid/.style={darkcandy, opacity=0.5},
]
\input{figure3b.tex}
\end{axis}
\end{tikzpicture}\hfill
\begin{tikzpicture}
\begin{axis}[
   width=2cm,
   height=1.5cm,
   scale only axis,
   xtick={1,2,3,4,5,6},
   xticklabels={$t_0$,$t_1$,$t_2$,$t_3$, $t_4$, $t_5$},
   ylabel={$\text{d}_B(\mathcal{P}^{t}_{\Rips},\mathcal{P}^{t+1}_{\Rips})$},
   ylabel near ticks,
   grid = none,
   minor tick num=3,
   every major grid/.style={darkcandy, opacity=0.5},
   ytick=\empty,
   axis lines*=left,
   cycle list name= color list
]
\input{figure3c.tex}
\end{axis}
\end{tikzpicture}\hfill
\begin{tikzpicture}
\begin{axis}[
   width=2cm,
   height=1.5cm,
   scale only axis,
   xtick={1,2,3,4,5,6},
   xticklabels={$t_0$,$t_1$,$t_2$,$t_3$, $t_4$, $t_5$},
   ylabel={$d_{W_1}(\mathcal{P}^{t}_{\Rips},\mathcal{P}^{t+1}_{\Rips})$},
   ylabel near ticks,
   ytick=\empty,
   grid = none,
   minor tick num=3,
   every major grid/.style={darkcandy, opacity=0.5},
   axis lines*=left,
   cycle list name= color list
]
\input{figure3d.tex}
\end{axis}
\end{tikzpicture}\hfill
\begin{tikzpicture}
\begin{axis}[
   width=2cm,
   height=1.5cm,
   scale only axis,
   xtick={1,2,3,4,5,6},
   xticklabels={$t_0$,$t_1$,$t_2$,$t_3$, $t_4$, $t_5$},
   ylabel={$d_B(\mathcal{P}^{t}_{\Wit},\mathcal{P}^{t+1}_{\Wit})$},
   ylabel near ticks,
   ytick=\empty,
   grid = none,
   minor tick num=3,
   every major grid/.style={darkcandy, opacity=0.5},
   axis lines*=left,
   cycle list name= color list
]
\input{figure3e.tex}
\addplot+[black, thick, dashed, name path=B] coordinates {(1,1500)(2,1500)(3,1500)(4,1500)(5,1500)(6,1500)} node [pos=0.8, fill=white, below right] {\color{black}$\infty$};
\addplot+[draw=none,name path=B] coordinates {(1,0)(2,0)(3,0)(4,0)(5,0)(6,0)};
\end{axis}
\end{tikzpicture}\hfill
\begin{tikzpicture}
\begin{axis}[
   width=2cm,
   height=1.5cm,
   scale only axis,
   xtick={1,2,3,4,5,6},
   xticklabels={$t_0$,$t_1$,$t_2$,$t_3$, $t_4$, $t_5$},
   ylabel={$d_{W_1}(\mathcal{P}^{t}_{\Wit},\mathcal{P}^{t+1}_{\Wit})$},
   ylabel near ticks,
   ytick=\empty,
   grid = none,
   minor tick num=3,
   every major grid/.style={darkcandy, opacity=0.5},
   axis lines*=left,
   cycle list name= color list
]
\input{figure3f.tex}
\end{axis}
\end{tikzpicture}
\begin{tikzpicture}
\begin{axis}[
  width=2cm,
  height=1.5cm,
  scale only axis,
  axis lines=left,
  axis line style={black},
  x tick label style={/pgf/number format/1000 sep=},
  enlargelimits=0.1,
  legend style={
  at={(1.3,0.4)},
  anchor=west,
  draw=none,
  legend columns=2},
  mark size=1pt,
  xtick={1,2,3,4,5,6},
  ytick={0.2,0.8},
  xticklabels={$t_0$,$t_1$,$t_2$,$t_3$, $t_4$, $t_5$},
  legend entries={
                  $\mu_X$,
                  $\tilde{\mu}_X$,
                  $\sigma_X$,
                  $\tilde{\sigma}_X$,
                  $\frac{\sigma_X}{\mu_X}$,
                  $\frac{\tilde{\sigma}_X}{\tilde{\mu}_X}$,
                  $w_1$,
                  $\tilde{w}_1$
                  }
]
\addplot+[lightcandy, mark=*,mark options={fill=lightcandy}, name path = A]
  coordinates {(1,0.47674345652) (2,0.47113643671) (3,0.49324005108) (4,0.55145091018) (5,0.66325900145) (6,0.66354515531)};
\addplot+[darkcandy, mark=*,mark options={fill=darkcandy}, name path = B]
  coordinates {(1,0.47674345652) (2,0.46940951352) (3,0.54412739748) (4,0.60964375921) (5,0.69168695173) (6,0.69141828315)};
\tikzfillbetween[of=A and B]{lightcandy, opacity=0.1};

\addplot+[darkblue, mark options={fill=darkblue}, name path = C]
  coordinates {(1,0.54345721917324) (2,0.57432907917) (3,0.61705207247) (4,0.62769015889) (5,0.70765198197) (6,0.76579081942)};
\addplot+[lightblue, mark options={fill=lightblue}, name path = D]
  coordinates {(1,0.54345721917) (2,0.6619320595) (3,0.72084605177) (4,0.73278177799) (5,0.88376578507) (6,0.97348897936)};
\tikzfillbetween[of=C and D]{darkblue, opacity=0.1};

\addplot+[darkgold, mark options={fill=lightgold}, name path = E]
  coordinates {(1,0.83014734374) (2,0.83006981543) (3,0.82656304895) (4,0.82549085249) (5,0.81767813098) (6,0.80872008382)};
\addplot+[lightgold, mark options={fill=darkgold}, name path = F]
  coordinates {(1,0.83014734374) (2,0.8294138344) (3,0.82359613361) (4,0.82219526757) (5,0.81153011173) (6,0.79843199462)};
\tikzfillbetween[of=E and F]{lightgold, opacity=0.1};

\addplot+[gray, mark options={fill=gray}, name path = G]
  coordinates {(1,0.29684777643) (2,0.29639759279) (3,0.3249275297) (4,0.33224579279) (5,0.37080154515) (6,0.41272653753)};
\addplot+[black, mark options={fill=black}, name path = H]
  coordinates {(1,0.29684777643) (2,0.29584317733) (3,0.3789881125) (4,0.32746485752) (5,0.39234811315) (6,0.44149606376)};
\tikzfillbetween[of=G and H]{gray, opacity=0.1};
\end{axis}
\end{tikzpicture}
\caption{Bottleneck and $L^1$-Wasserstein distances between the persistence diagrams in iteration $t$ and $t+1$. The persistent homology has been computed on the Delaunay complex, Vietoris-Rips complex and the witness complex, respectively. A total of $250$ samples from a signature collection are represented \cite{MOBISIG}. Each line corresponds to a single sample and the lines are colored corresponding to one of six selected users in \textbf{\color{gray}$\bullet$} gray, \textbf{\color{black}$\bullet$} black, \textbf{\color{lightblue}$\bullet$} blue, \textbf{\color{lightgold}$\bullet$} yellow, \textbf{\color{orange}$\bullet$} orange and \textbf{\color{darkcandy}$\bullet$} red.}
\label{fig:stopping}
\end{figure}

All source code is written in Python 3.7. The GUDHI \cite{gudhi:urm} library is used for the calculation of simplicial complexes, filtrations and persistent homology. We investigate $83$ users, considering $45$ signatures per user from the MOBISIG signature database \cite{MOBISIG}, which show the same letters, but are independent writings. For each user we have a set of $45$ persistence diagrams and a set of $45$ corresponding handwritings. In every iteration as many new points are added as are already in the respective example of a signature. We inserted the points uniformly within the convex hull of the initial point set, see Fig. \ref{fig:voronoi}.

\subsection{Experimental Setting}
$\Rips(X)$ is expanded up to the third dimension. The \textit{maximum edge length} is set to the average edge length between two points within the data set. We use the same $r$ for $\Cech(X,r)$ and $\Rips(X,r)$, so that $\Rips(X,r)$ differs topologically more from the union of closed balls around each point, but is faster to compute. Finding an optimal radius as distance threshold is considered open \cite{ZomorodianC05}, thus we use $r = \max_{x,y \in X} ||x-y||$ as empirical heuristic.

The strong $\Wit(X,r)$, embedded into $\mathbb{R}^d$, is recalculated for each sample at each interpolation step. We select uniformly $5\%$ of the points as landmarks. We set $\alpha = 0.01, \gamma = 0.1$ and $p = 1$.

We assume that the persistence diagrams are i.i.d. A free parameter $\alpha$ quantifies a tolerance to topological change, thus a decision must be made on the following hypotheses about the distributions of the persistence diagrams:
\begin{center}
  \begin{itemize}
    \item[(a)] $\text{H}_0: \mathcal{P}_K^t$ and $\mathcal{P}_K^{t+1}$ have different underlying distributions and
    \item[(b)] $\text{H}_1: \mathcal{P}_K^t$ and $\mathcal{P}_K^{t+1}$ have the same underlying distribution.
  \end{itemize}
\end{center}
We use an asymptotic solution for testing by trimmed Wasserstein distance \cite{czado1998assessing}:
\begin{equation}
\hat{\Gamma}^p_\gamma(\mathcal{P}^t_K,\mathcal{P}^{t+1}_K) = \frac{1}{(1-2\gamma)} \left( \sum_{j=1}^{m} ||(\mathcal{P}^t_K)_j-\mathcal{P}^{t+1}_K)_j||^p_\infty \Delta \gamma \right)^{1/p}.
\end{equation}
The trimming bound $\alpha \in [0, 1/2)$ results from the integral for the continuous case as a difference in a finite weighted sum. It is computed using the expected value of the persistence diagrams and is exact in the limit $\int_\gamma^{1-\gamma} f(x) dx = \lim_{\Delta\gamma \rightarrow 0} \sum_{x \in X} f(x) \Delta \gamma$. The critical region for our hypothesis $\text{H}_0$ against $\text{H}_1$ is
\begin{equation}
\left(\frac{nm}{n+m}\right)^{\frac{1}{p}} \frac{\hat{\Gamma}^p_\gamma - \alpha^p}{\hat{\sigma}_\gamma} \leq z_\gamma,
\end{equation}
where $z_\gamma$ denotes the $\gamma$-quantile of the standard normal distribution and $n=m$, with $n$ being the number of samples. The initial problem can be rephrased as
\begin{center}
  \begin{itemize}
    \item[(a)] $\text{H}_0: \Gamma_\gamma^p (\mathcal{P}^t_K,\mathcal{P}^{t+1}_K) > \alpha$ and
    \item[(b)] $\text{H}_1: \Gamma_\gamma^p (\mathcal{P}^t_K,\mathcal{P}^{t+1}_K) \leq \alpha$.
  \end{itemize}
\end{center}

\subsection{Evaluation}
In Fig. \ref{fig:stopping}, seventh diagram, elementary statistics are computed for the entire data set such as mean $\mu_X$, standard deviation $\sigma_X$, variation $\frac{\sigma_X}{\mu_X}$ and $d_{W_1} = d_{W_1}(X_{\text{org}},X^{t})$. The statistics are also computed for the interpolated data with topological stop, respectively, marked with $\sim$.

We achieved an improvement for each measured statistic at each iteration step using topological stop. In Fig. \ref{fig:stopping} the topological similarity between the individual users are made visual. $\Rips(X,r)$ and $\Del(X,r)$ seem suitable to estimate the homology groups, whereas $\Wit(L,W)$ produced far less stable results, due to the small selection of landmarks.

\section{Conclusions}
\label{sec:conclusion}
We have discussed the connection of Voronoi diagrams to the Delaunay complex and its connection to other complexes, which should serve as a basis to explore related algorithms to the Voronoi interpolation. We investigated into metrics to measure differences in persistent homology and could visualize the changing homology groups of the users signatures during interpolation, see Fig. \ref{fig:stopping}.
Our result is a stopping-criterion with a hypothesis test to determine whether the persistent homology of an interpolated signature still originates from the same distribution as the source. Our measurements show an improvement of statistics compared to vanilla Voronoi interpolation. We demonstrated, that -- under mild conditions -- the Delaunay complex, \v{C}ech-Delaunay complex and Wrap complex can also be used for filtration up to homotopy equivalence. Following open research questions arose during our investigations:
\begin{itemize}
  \item The intrinsic geometry of the data points is often not the Euclidean one. On the other hand side the frequently used embedding of the Voronoi tessellation is. This causes unwanted artifacts. Is there a geometrically meaningful clipping for general metric spaces, for example using geodesics in a smooth manifold setting? In which manifold should $\Del(X)$ be embedded?
  \item To our knowledge there is no evidence known that the Voronoi tessellation obtains the homology groups. According to \cite{reem2011geometric}, the Voronoi tessellation is stable. However, the experiments show that for increasing iterations additional homology groups appear. Does the Voronoi tessellation preserve homology groups and homotopy groups in general metric spaces?
\end{itemize}

\subsubsection*{Acknowledgements.} We would like to thank David Haller and Lekshmi Beena Gopalakrishnan Nair for proofreading. Further, we thank Anton Rechenauer, Jan Frahm and Justin Noel for suggestions and ideas on the elaboration of the final version of this work. We express our gratitude to Demian Vöhringer and Melanie Sigl for pointing out related work. This work was partially supported by Siemens Gas and Power GmbH \& Co. KG.

\subsubsection*{Code \& Data.} The implementation of the methods, the data sets and experimental results can be found at: \url{https://codeberg.org/Jiren/SIML}.
%
%
\bibliographystyle{splncs03}
\bibliography{paper5}
\end{document}